\documentstyle[aas2pp4]{article}

\def\ie{ i.e.}
\def\msun{$M_\odot$}

\def\fun#1#2{\lower3.6pt\vbox{\baselineskip0pt\lineskip.9pt
  \ialign{$\mathsurround=0pt#1\hfil##\hfil$\crcr#2\crcr\sim\crcr}}}
\def\la{\mathrel{\mathpalette\fun <}}
\def\ga{\mathrel{\mathpalette\fun >}}
\def\eg{{ e.g., }}
\def\etal{{et al. }}

\def\he#1{\hbox{${}^{#1}{\rm He}$}}

\def\bline{\rule[1.2mm]{3em}{0.1mm}}

\newlength{\headroom}
\newlength{\psfigskip}
\begin{document}
\input epsf
\def\msun{M_\odot}
\def\la{\mathrel{\mathpalette\fun <}}
\def\ga{\mathrel{\mathpalette\fun >}}
\def\fun#1#2{\lower3.6pt\vbox{\baselineskip0pt\lineskip.9pt
  \ialign{$\mathsurround=0pt#1\hfil##\hfil$\crcr#2\crcr\sim\crcr}}}
\def\3he{$^3$He}
\def\4he{$^4$He}
\def\6li{$^6$Li}
\def\7li{$^7$Li}
\def\he3{$^3$He}
\def\eg{{ e.g.}}
\def\ie{{ i.e.}}
\def\etal{{ et al.~}}
\def\hii{H\thinspace{$\scriptstyle{\rm II}$}~}
\def\popii{Pop\thinspace{$\scriptstyle{\rm II}$}~}
\def\Yp{Y$_{\rm p}$}
\def\adot{\odot{a}}
\lefthead{Graff et al.}
\righthead{Constraining stellar remnants with multi-TeV $\gamma$-rays}
\title{Constraining the cosmic abundance of stellar remnants \\
 with multi-TeV $\gamma$-rays}

\author{David~S.~Graff\altaffilmark{1,2}, Katherine~Freese\altaffilmark{3,4}, Terry~P.~Walker\altaffilmark{1,2}, and Marc~H.~Pinsonneault\altaffilmark{2}}
\authoremail{graff.25,walker.33@osu.edu; ktfreese@umich.edu; pinsono@astronomy.ohio-state.edu}
\altaffiltext{1}{Department of Physics, The Ohio State University, Columbus, OH 43210}
\altaffiltext{2}{Department of Astronomy, The Ohio State University, Columbus, OH 43210}
\altaffiltext{3}{Department of Physics, University of Michigan, Ann Arbor, MI
48109}
\altaffiltext{4}{Visitor, Max Planck Institut f\"{u}r Physik, F\"{o}hringer Ring 6, M\"{u}nchen, Germany}

%

\begin{abstract}
If galactic halos contain stellar remnants, the infra-red flux from
the remnant progenitors would contribute to the opacity of multi-TeV
$\gamma$-rays.  The multi-TeV $\gamma$-ray horizon is established to
be at a redshift $z>0.034$ by the observation of the blazar Mkn501 .
By requiring that the optical depth due to $\gamma \gamma
\longrightarrow e^+ e^-$ be less than one for a source at $z=0.034$ we
limit the cosmological density of stellar remnants, $\Omega_{rm} \le
(2-4) \times 10^{-3} h_{70}^{-1}$ ($h_{70}$ is the Hubble constant in
units of 70 km sec$^{-1}$ Mpc$^{-1}$) and thus strongly constrain
stellar remnants as a cosmologically significant source of dark
matter.
\end{abstract}

%

\setcounter{footnote}{0}
\twocolumn

\section{Introduction}

The constituents of the dark matter observed in galactic halos are, to
date, unknown.  Gravitational microlensing experiments, MACHO and
EROS, have detected several unexplained events in the direction of the
Magellanic Clouds ({\sc Macho}s) (Alcock\etal \cite{macho:2yr},
Renault \etal \cite{ren}, Palanque-Delabrouille -etal
\cite{erossmc1}).  Their combined results limit Macho masses in the
range $(10^{-7} - 0.02)$M$_\odot$ to be less than 20\% of the Halo
(Alcock \etal \cite{erosmacho}).  A standard interpretation of
the microlensing results is that perhaps 30\% of the galactic halo is
composed of objects of mass roughly $\sim 0.5 \msun$ (Alcock \etal
\cite{macho2yr}), but this interpretation has several problems: Hubble
Deep Field star counts (Bahcall \etal \cite{bahcall},Graff \& Freese
\cite{gf96a}) and an extrapolation of parallax data (Dahn \etal
\cite{dahn}) suggest that faint stars (0.08$\msun$-0.2$\msun$) and
brown dwarfs ($\la 0.08$ M$_\odot$) contribute negligibly to the
Galactic halo ($<$1\% of the Halo mass)(Graff \& Freese \cite{gf96b};
Mera, Chabrier \& Schaeffer \cite{mcs96}; Gould, Flynn \& Bahcall
1998).  Gyuk, Evans \& Gates \cite{geg} have also argued that the {\sc
Macho} lenses cannot be a Halo or Spheroid population of brown dwarfs.

Stellar remnants might make possible {\sc Macho} candidates; they have the right mass and are dark, but they also have their problems:  Fields, Freese, and Graff (\cite{ffg}) extrapolated the
Galactic population of {\sc Macho}s to cosmic scales and found a mass
density that is comparable to the cosmic baryon density implied by Big
Bang Nucleosynthesis; this result is independent of the nature of the
{\sc Macho}s.  If the {\sc Macho}s are stellar remnants, then the
additional mass density of the gas left over from the progenitors
becomes a problem: virtually all of the baryons of the universe have
to be recycled through the {\sc Macho}s and their progenitors.
Over-pollution is difficult to avoid if these stellar remnants make up
a significant fraction of the dark matter in galactic halos.  It is
problematic to hide the gas and/or metals ejected during the formation
of the remnants (Gibson and Mould \cite{gm}; Fields \etal \cite{ffg}).
If carbon and nitrogen do not leave the stars, as suggested by
Chabrier (\cite{chab99}), then the pollution is less problematic,
although helium is still very restrictive (Fields, Freese, and Graff
\cite{ffg2}).  These problems of baryonic mass budget and chemical
overproduction are particularly severe for higher mass progenitors
that give rise to neutron stars (Venkatesan, Olinto \& Truran
\cite{vot}).

In this {\sl Letter} we constrain stellar remnant baryonic dark matter
by examining the infra-red radiation that is emitted during the
evolution of the remnants' progenitors while on the main-sequence and
during the subsequent red giant phase.  This diffuse infra-red
background (DIRB) contributes to the opacity of multi-TeV
$\gamma$-rays allowing observations of TeV gamma ray sources to limit
the DIRB (Gould \& Schr\'{e}der \cite{gs66}, Stecker, De Jager \&
Salamon \cite{stecker}, MacMinn \& Primack \cite{macminn}).  HEGRA CT1
detections of multi-TeV $\gamma$-rays from the blazar Mkn501 (Bradbury
\etal \cite{hegra}) suggests that the universe is optically thin to 10
TeV $\gamma$-rays out to $z=0.034$ and thus limits the DIRB (Funk
\etal \cite{funk98}).  We use this multi-TeV $\gamma$-ray horizon to
constrain the DIRB expected from remnant halos.  We show that the
contribution of remnants to baryonic halo dark matter must be
$\Omega_{\rm rm} h_{70} < 4 \times 10^{-3}$ where $h_{70}$ is the
Hubble constant in units of 70 km s$^{-1}$Mpc$^{-1}$ and $\Omega_{\rm
rm}$ is the density of stellar remnants in units of the critical
density $\rho_c=3H^2/8\pi G$.  Note that the existing constraints on
the DIRB obtained by measuring the infrared background directly with
DIRBE (Hauser 1995), IRAS (Boulanger and Perault 1988), and other
experiments are weaker than the constraints obtained by HEGRA for the
energy range of interest.

\section{Measuring the DIRB with multi-Tev $\gamma$-rays}

The Universe appears to be optically thin to multi-TeV $\gamma$-rays
out to a redshift of (at least) $z\sim 0.03$ based on the detection of
two blazars, Mkn421 ($z=0.031$) and Mkn501 ($z=0.034$).  Neglecting
self-absorption, the dominant source of opacity for multi-TeV
$\gamma$-rays (energy $E(z) = (1+z)E$) is pair-creation off diffuse
background photons (energy $\epsilon(z) = (1+z)\epsilon$) which has a
threshold of

\begin{eqnarray}
\epsilon_{\rm Th} & = & \frac{2m_e^2}{E(1-\cos \theta)(1+z)^2} \nonumber \\
                  & = & \frac{0.5 }{E_{\rm TeV}(1-\cos \theta)(1+z)^2}{\rm eV},
\end{eqnarray}
where $\theta$ is the relative photon scattering angle in the
rest frame of the microwave background, $E_{\rm TeV}$ is the observed
source photon energy measured in TeV, and we 
have fixed $\hbar = k = c = 1$.  The $\gamma\gamma\longrightarrow
e^+e^-$ cross section is Bethe-Heitler and is plotted in Figure 1:
\begin{eqnarray}
{\sigma_{\gamma\gamma}(E(z),\epsilon(z),\theta)\,  = }\hspace{1.5in} \nonumber\\
{\frac{3\sigma_T}{16}(1-\beta^2)\left[2\beta(\beta^2-2)+(3-\beta^4)\ln\left(
\frac{1+\beta}{1-\beta}\right)\right]},
\end{eqnarray}
where $\beta^2 \equiv 1-(\epsilon_{Th}/\epsilon)^2$ and $\sigma_T = 6.65\times
10^{-25} {\rm cm}^2$ is the Thomson cross section.  

\begin{figure}[htb]
\plotone{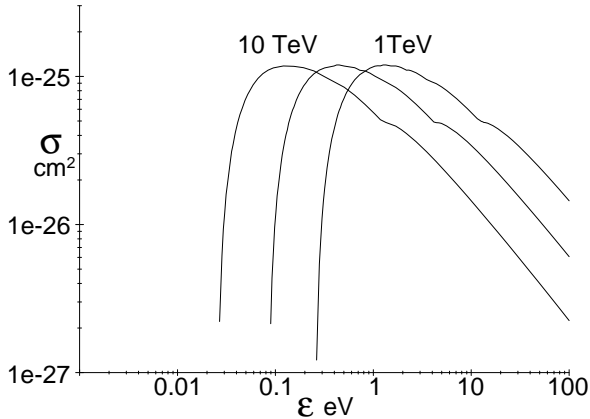}
\caption[Fig1]{\label{fig1} The cross section for $\gamma\gamma
\longrightarrow e^+ e^-$.  From left to right,
the three curves correspond to three different Blazar photon energies,
10, 3 \& 1 TeV.  We can place limits on the DIRB where the cross
section is large, in the range $0.03-3$ eV. 
The blips around 10 ev are numerical artifacts.}
\end{figure}

As discussed below, we will be able to
place limits on the DIRB in the energy range where the cross
section is large, $0.03-3$ eV.  The optical depth $\tau_{\gamma
\gamma}$ at observed source energy $E$ out to redshift $z$ due to a comoving 
background spectrum $(1+z)^3n(\epsilon)d\epsilon$ is
\pagebreak
\begin{eqnarray}
\label{tau}
{\tau_{\gamma\gamma}(E,z) =} \hspace{4cm}\nonumber\\
 \int_{0}^{z} \!\! dz\frac{dx}{dz}\int_{-1}^{1} \!\! d\mu
\frac{1-\mu}{2}\int_{\epsilon_{\rm Th}}^{\infty} \!\!\! d\epsilon
(1+z)^3 n(\epsilon)\sigma_{\gamma\gamma},
\end{eqnarray}
where $\mu=\cos\theta$ and for a flat Universe, $(1+z)dx/dz = H^{-1} = H_0^{-1}[\Omega(1+z)^3+
\Omega_\Lambda]^{-1/2}$
with $H$ the Hubble parameter, $\Omega$ the present matter density
and $\Omega_\Lambda$ the present cosmological constant energy density
in units of $\rho_c$.
Note that for sources at z=0.03, the value of $\Omega_\Lambda$ is irrelevant
to the results.

\section{The DIRB Produced by Remnant Halos}


In order to determine the DIRB produced by the progenitor stars of
remnant {\sc Macho}s, we start with stellar models of intermediate
mass, $(2-9)\msun$, and low metallicity, $Z=10^{-4}$, generated
specifically for this project.  The models use the nuclear reaction
cross-sections of Bahcall, Bau \& Pinsonneault (1998).  The equation
of state is fully ionized in the interior, and the model uses the Saha equation
at low temperatures (Guenther, Demarque, Kim \& Pinsonneault 1992).
We used the model opacities of Alexander \& Ferguson (1994) for
$T<10,000 K$, and OPAL opacities for higher temperatures (Iglesias \&
Rogers 1996).  We assumed grey atmospheres, as is appropriate for
these high temperature models.  Our models ran from 
zero age main sequence (ZAMS) to He core exhaustion. 

For comparison, Girardi \etal (1996) have also produced models
for low Z isochrones, and find results that are very similar;
the similarity suggests that our models are 
theoretically robust.  Girardi \etal claim that the post He core
burning lifetime of the star is of order 0.3\% of the main sequence
lifetime, and can thus be ignored here.  Our models contain no
convective overshoot, and so conservatively underestimate the total
light emitted by stars.  Were we to follow the convective overshoot
system of Girardi \etal (1996), we would expect stars to live
aproximately 20\% longer, emit a total of 20\% more light, and thus
our limits would be 20\% more restrictive.

In addition to the $Z=10^{-4}$ models discussed above, we also
examined models with $Z=10^{-8}$ and with zero metallicity.  We found
no substantive changes in the stellar models, with the following
exception: for progenitors at the high mass end (9$\msun$), stars of
the {\it extremely} low metallicity of $Z=10^{-8}$ gave similar results
to the low $(Z=10^{-4})$ models; however, stars of strictly zero
metallicity actually never evolved off the main sequence and produced
about half as many infra-red photons as the low metallicity models.
Although our limits would then be only half as severe, such a scenario
of all of the stars having not even the merest whiff of metallicity is
probably academic.

Note that the mass range of our stellar models, $(2-9) M_\odot$,
covers all the allowed white dwarf progenitors as well as some low
mass neutron star progenitors.  Remnant masses are taken from the
models of Van den Hoek \& Groenewegen (\cite{vanden}).  Stars below
$2\msun$ would leave remnants so bright that they would have been
detected already (Graff, Laughlin \& Freese \cite{glf}).  The stellar
models for stars above $9\msun$ are not considered as reliable as
those for lower mass stars.  In addition, as mentioned in the
Introduction, Fields \etal (1998) showed that the mass budget and
chemical abundance problems are particularly severe for the highest
mass progenitors that give rise to neutron stars.

We model the spectrum of light emitted by a star at a particular stage
of stellar evolution as a black body spectral shape characterized by
the effective temperature of the star.  Unless there are features in
the stellar photon spectrum that are much broader than the width of
the cross section (as a function of infrared photon energy; the width
$\sim \epsilon_{\rm th}$), the optical depth is fairly insensitive to
the spectral shape and is largely determined by the total number of
photons with energy above $\epsilon_{\rm th}$ (see Fig. 1).  Any
spectral features, even broad absorption bands, will have only a
minimal effect on the optical depth.  If anything, realistic spectra
would show absorption at the ultraviolet end and re-emission of the
same energy as more photons at the infrared end (see further
discussion of this point below).  Thus, a black body spectrum
conservatively underestimates the number of photons produced by a star
\footnote{An exception to this rule is low temperature zero
metallicity stars which emit more energy at high frequencies than
black body.  However, our stars are at too high a temperature for this
to be an issue.}.

Stars emit almost all their total energy during two distinct stages of
evolution, main sequence (MS) and Helium core burning (HB).  All other
stages of stellar evolution are much too short lived or too dim to be
significant in the total energy budget.  Thus, we approximate the
total energy emitted by a star as the sum of two black bodies
each marked by an average effective temperature $\langle T \rangle$
and a total emitted
energy $E_{\rm tot}$.  These two quantities 
are determined by integrating over the
stellar models as follows:
\begin{eqnarray}
{E_{\rm tot}}     & = & \int_{\rm stage} \!\!\!\!\! dt\,L(t) \\
\langle T\rangle & = & E_{\rm tot}^{-1} \, \int_{\rm stage} \!\!\!\!\! dt\, L(t) \, T(t) 
\, ,
\end{eqnarray}
where $T(t)$ is the effective temperature as a function of time, and $L(t)$ the luminosity.
We summarize the stellar
parameters adopted in Table 1, where we have given the initial
mass of the progenitor, the remnant mass, the average effective
temperatures of the main sequence $\langle T \rangle_{\rm MS}$ 
and Helium core burning phases $\langle T \rangle_{\rm HB}$,
and the log of the corresponding total emitted energies of the two phases.

\vspace{0.5cm}
\noindent
\small
\begin{tabular}[!t]{|cccccc|}
\hline
Initial       & Remnant      & $\langle T\rangle_{\rm MS}$ & $\log(E_{\rm MS})$ & $\langle T\rangle_{\rm HB}$ & $\log(E_{\rm HB})$ \\
$(\msun)$  & $(\msun)$ &  $(^\circ$K)                & (ergs)            &   $(^\circ$K)              & (ergs)       \\ \hline   
2             & 0.68         & 13260                     & 63.30           &  5424                    & 62.65 \\
4             & 0.91         & 19840                     & 63.59           &  12880                   & 63.22 \\
9             & 1.84         & 28940                     & 64.04           &  19290                   & 63.49 \\ \hline
\end{tabular}
\normalsize

{Table 1: Stellar parameters from our model}
\vspace{0.5cm}

The total number of photons emitted by a single star at energy
$\epsilon$ can thus be taken as the sum of the number produced by
blackbody emission by the MS and RG phases,
\begin{eqnarray}
{N(\epsilon)\, d\epsilon = } \hspace{4cm}\nonumber \\
 \sum_{i=\{MS, RG\}} \!\!\!\! E_{{\rm tot},i}
 \frac{15}{(\pi\,k_b\,\langle T \rangle_i)^4}
\frac {\epsilon^2}{\exp(\epsilon/k_b \langle T \rangle_i) - 1} d\epsilon
\, .
\end{eqnarray}
Note that this equation has been normalized by requiring that
$\int N(\epsilon)_i \epsilon d\epsilon = E_{\rm tot,i}.$
We then integrate over the redshifts at which the radiation is emitted.
The co-moving number density of white dwarfs is
$\Omega_{\rm rm} \rho_c m_{\rm rm}^{-1}$.
Thus we find that the co-moving number density of 
background photons with present day energy $\epsilon$ is
\begin{equation}
\label{n}
n(\epsilon) = \int dz\, \Omega_{\rm rm} \rho_c m_{\rm rm}^{-1}
N(\epsilon(1+z)) \, .
\end{equation}

\section{Results}

Since there is no observed break in the HEGRA gamma ray spectrum from
Mkn 501 at redshift z=0.034 (Funk \etal \cite{funk98}), we can
conservatively conclude that $\tau_{\gamma\gamma}<1$ out to z=0.034.
Using equations (\ref{tau})
and (\ref{n}), we can then constrain the mass density of white dwarfs
due to the infrared emission by their progenitors.  As mentioned
previously, we consider a variety of progenitor masses in the range
(2-9)$\msun$.  Inspection of Fig. 1 shows that we can obtain
upper limits on the infrared photon density over the
energy range $3 \times 10^{-2} - 3$ eV, where the cross section
for interaction is sizable (this range would be extended
for a hard spectrum).  For each progenitor mass, we vary the gamma ray energy
in the observed range (1-10) TeV to find the largest resultant value
of optical depth $\tau_{\gamma\gamma}$.  [Note that the gamma ray
energy $E$ is inversely proportional to the range of affected infrared
photons (c.f. Eq. (1)).] 

We also consider a variety of redshifts at
which the stellar light is emitted.  
Our limit in Eqn. (8) below applies to progenitor stars created 
at $z<60$, but with some decreased sensitivity to stars in the
mass range (3-4)$\msun$ created at $z>15$.  
Stars absorb photons above the Balmer cutoff at 0.5 eV,
so that these photons are not available to scatter with the TeV
gamma rays from Mkn 501.
At $z=15$, the Balmer cuttoff at 0.5
eV passes the threshold of a 10TeV photon at 0.03 eV;
thus no light emerges from the star that can scatter with the 
TeV gamma rays.  Since 2$\msun$
stars emit most of their light just to the red of the Balmer cuttoff, we
will not see the light from these stars emitted at $z>15$.  However, a
low metallicity 2$\msun$ star will live for 600 Myr 
(assuming no convective overshoot), 
whereas the universe at $z=15$ is only
150$h_{70}^{-1}$ Myr old.  Thus, even if
a 2$\msun$ star is created at very high redshifts, 
it will emit most of its light
after $z=15$.  A $(4-9)\msun$ mass star will be hot enough in its main
sequence phase to emit much light at energies higher than the
Balmer cutoff; we will see light from that star out to z=60,
the redshift at which the Lyman cutoff is redshifted below
0.03 eV.  There is still a small hole of decreased sensitivity to stars
in the mass range $3 - 4 \msun$ created at $15<z<60$ because of
the absorption of light blueward of the Balmer cutoff; these
stars are shorter lived than the 2$\msun$ stars and do emit a significant
amount of their light within 150 Myr of their formation.
We do not expect any significant amount of baryonic objects to form
before $z=60$ (Tegmark \etal \cite{tsrbap}).

Combining equations (\ref{tau}) and (\ref{n}), we find a
robust limit of
\begin{equation}
\label{limit}
\Omega_{rm} h_{70} < (2-4) \times 10^{-3} \, .
\end{equation}

Since our limit depends on the number density of photons, which does
not vary with redshift, and since we are sensitive to a large range of
energies, our limit on $\Omega_{rm}$ 
is not sensitive to the redshift at 
which the stellar light is emitted (except for the decreased
sensitivity discussed above for $z>15$ and (3-4)$\msun$).
In addition, the limit is relatively insensitive to the 
star formation rate and to the initial
mass of the progenitor star (with stellar mass between 2 and 9 $\msun$).
The limits quoted above represent the full range
of limits found for any star formation rate, initial mass function,
and formation redshift within the broad bounds described above.
The key variables that govern the severity of the limits are:
higher mass stars give slightly more severe limits, and a ``burst''
star formation rate, which concentrates the emitted light within a
narrow range of redshifts $(\delta z/z \le 1)$ will give more severe
limits (for the burst, no photons redshift out of the testable energy
range).

One possible way to escape this limit is dust.  It is possible that
the progenitor stars are so enshrouded by dust that little of their
light escapes, and if the dust is cool enough, that the light is
re-radiated away at too long a wavelength to interact with the TeV
$\gamma$-rays.  Such a situation is most likely at high redshifts,
where the dust could absorb the UV photons that later (by z=0.034)
would have been redshifted into the infrared (IR).
However, at lower redshifts ($z<10$), dust could actually make
the limits stronger: dust could absorb UV photons and
reradiate them in the IR, causing much more
absorption in the DIRB.
Dust absorbtion is very model dependent, depending on the type of
dust, the dust geometry and temperature, and beyond the scope of this
work.

\section{Interpretation and comparison with other limits}

Our result constrains the cosmological abundance of stellar remnants.
The microlensing experiments have only detected {\sc Macho}s around
a single galaxy, the Milky Way.  However, if we assume that the Milky
Way is not a special galaxy, and that other, similar galaxies also
have their coterie of Halo {\sc Macho}s, then the Universe should be filled
with {\sc Macho}s.

Dalcanton \etal (\cite{dalcanton}) searched directly for 
a cosmological population of {\sc Macho}s by looking for a signal of
amplification of continuum emission of QSO's at high redshift.  The
fact that they did not find such an amplification allowed them to
constrain all compact objects (not just remnants) in the mass range
$(0.1 M_\odot - 10 M_\odot)$ to $\Omega_{\rm Macho} \leq 0.1$.

Fields \etal (1998) examined the cosmic abundance of
{\sc Macho}s.  They found that a simple extrapolation of the (supposed)
Galactic population of {\sc Macho}s to cosmic scales gives a cosmic density
\begin{equation}
\label{cosmic}
\Omega_{\rm Macho} = (0.0051-0.024)f_{gal} h_{70}^{-1} \, . 
\end{equation}
Here $f_{gal}$ is the fraction of
galaxies that contain {\sc Macho}s.  To obtain an estimate of a lower limit
to $f_{gal}$ under reasonable assumptions, they considered the limiting case where only galaxies
within one magnitude of the Milky Way (i.e., $M_{MW} \pm 1 {\rm mag})$
contain {\sc Macho}s, and only spiral galaxies contain {\sc Macho}s, and found
$f_{gal} > 0.17$.
We consider these two results, the limit from Dalcanton \etal
and Eq. (\ref{cosmic}), to be the most robust since they apply
to all possible {\sc Macho}s, and not just stellar remnants.

Fields, Freese \& Graff (1998) also placed a much stronger limit on the
cosmological density of stellar remnants by noting that the early
universe had a low carbon and/or nitrogen enrichment.  Using low metallicity 
stellar yields of van den Hoek and Groenewegen (\cite{vanden}),
they found that only about 10$^{-2}$ of all baryons can have passed
through the white dwarf progenitors, i.e., $\Omega_{rm} < \Omega_{prog}
<10^{-2} \Omega_B < 4.5 \times 10^{-4} h_{70}^{-2}$, 
where $\Omega_B$ is the cosmic baryon density.
However, as noted by Chabrier (1999),
this limit depends on the carbon yields of zero metallicity intermediate
mass stars.

Our new limit in Eq. (\ref{limit}), based on the DIRB, is less general
than limits on all {\sc Macho}s, but more robust than the carbon limit
placed by Fields \etal (1998): although the carbon yields
may be uncertain, intermediate mass stars certainly do produce light.
Our new limit applies to remnants with any initial mass function
(between 2 and 9$\msun$) and any star formation rate 
(with a somewhat less restrictive limit for light emitted at
$15<z<60$ and for $2.5 - 4 \msun$), and is thus extremely model independent.
We can compare our new limit in Eq. (\ref{limit}) with the
extrapolation of the Milky Way abundance of {\sc Macho}s placed by
Fields \etal (1998) in Eq. (\ref{cosmic}).  If all galaxies
contain {\sc Macho}s in the same abundance as the Milky Way does,
then the Machos cannot be stellar
remnants ($h_{70}=1$).  Of course, it is impossible to place limits on
the nature of the Milky Way Halo based only on cosmological limits: the
Milky Way could be the only galaxy in the universe to have {\sc
Macho}s.  Still, based on this work, we can say that there must be
some galaxies whose halos are not dominated by white dwarfs.

\acknowledgements

We thank Julien Devriendt, Andy Gould and Leo Stodolsky for helpful
discussions and suggestions.  This work was supported at Ohio State by
DOE grant DE-AC02-76ER01545 and by DOE at the University of
Michigan. 

\pagebreak

\pagebreak

\end{document}